\documentclass[pra,twocolumn,floatfix,superscriptaddress]{revtex4}

\usepackage{graphicx}
\usepackage{bm}
\usepackage{amsmath,amssymb, amsthm}
\newcommand{\be}{\begin{equation}}
\newcommand{\ee}{\end{equation}}
\newcommand{\ba}{\begin{eqnarray}}
\newcommand{\ea}{\end{eqnarray}}

\begin{document}

\title{Decoherence in Adiabatic Quantum Computation}

\author{M.~H.~S.~Amin}
\affiliation{D-Wave Systems Inc., 100-4401 Still Creek Drive,
Burnaby, B.C., V5C 6G9, Canada}

\author{Dmitri V.~Averin}

\author{James A.~Nesteroff}

\affiliation{Department of Physics and Astronomy, Stony Brook
University, SUNY, Stony Brook, NY 11794-3800 }

\begin{abstract}

We have studied the decoherence properties of adiabatic quantum
computation (AQC) in the presence of in general non-Markovian, e.g.,
low-frequency, noise. The developed description of the incoherent
Landau-Zener transitions shows that the global AQC maintains its
properties even for decoherence larger than the minimum gap at the
anticrossing of the two lowest energy levels. The more efficient
local AQC, however, does not improve scaling of the computation time
with the number of qubits $n$ as in the decoherence-free case. The
scaling improvement requires phase coherence throughout the
computation, limiting the computation time and the problem size $n$.

\end{abstract}

\maketitle

The adiabatic ground-state scheme of quantum computation
\cite{farhi,gs} represents an important alternative to the
gate-model approach. In adiabatic quantum computation (AQC) the
Hamiltonian $H_S$ of the qubit register and its wave function
$|\psi\rangle$ undergo adiabatic evolution in such a way that, while
the transformations of $|\psi\rangle$ represent some meaningful
computation, this state also remains close to the instantaneous
ground state $|\psi_G\rangle$ of $H_S$ throughout the process. This
is achieved by starting the evolution from a sufficiently simple
initial Hamiltonian $H_i$, the ground state of which can be reached
directly (e.g., by energy relaxation), and evolving into a final
Hamiltonian $H_f$, whose ground state provides the solution to some
complex computation problem: $H_S {=} [1-s(t)]H_i + s(t)H_f$, where
$s(t)$ changes from 0 to 1 between some initial ($t_i{=}0$) and
final ($t_f$) times.

The advantage of performing a computation this way, besides its
insensitivity to gate errors, is that the energy gap between the
ground and excited states of the Hamiltonian $H_S$ ensures some
measure of protection against decoherence. This protection, as
partly demonstrated in this work, is not absolute. Nevertheless,
it allows for the ground state to maintain its coherence
properties in time far beyond what would be the single-qubit
decoherence time in the absence of the ground-state protection.
This feature of the AQC remains intact \cite{we1} even if the
decoherence strength and/or temperature is much larger than the
minimum gap.

\begin{figure}[t]
\includegraphics[width=6.2cm]{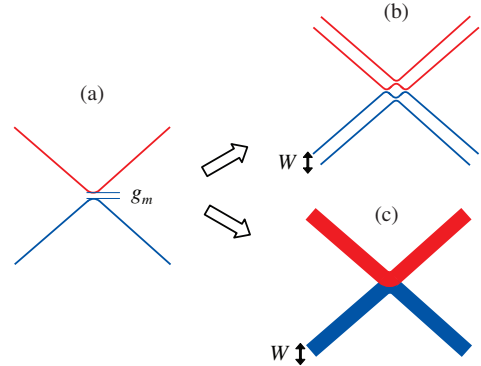}\\
\caption{Broadening of the energy levels of a closed system (a) due
to coupling to an environment made of (b) a single two-state system,
or (c) infinitely many degrees of freedom with a continuous energy
spectrum. In general, the coupling to an environment splits a single
anticrossing into $M^2$ anticrossings where $M$ is the number of
environment energy eigenstates. For environment with a continuous
spectrum the anticrossing turns into a continuous transition region
of width $W$.} \label{fig1}
\end{figure}

In general, the performance of an adiabatic algorithm depends on the
structure of the energy spectrum of its Hamiltonian $H_S$. Here we
consider a situation, which is typical for complex search and
optimization problems \cite{we1}, when the performance is limited by
the anticrossing of the two lowest energy states. The minimum gap
$g_m$ between those states shrinks with an increasing number $n$ of
qubits in the algorithm, although the exact scaling relation is not
known in general. In an isolated system with no decoherence, the
limitation is due to the usual Landau-Zener tunneling at the
anticrossing, which drives the system out of the ground state with
the probability given by the ``adiabatic theorem''. Different
formulations of the theorem all give the computation time as some
power of the minimum gap: $t_f \propto g_m^{-\delta}$
\cite{a_th,AdTheorem}.

The main assumption behind the adiabatic theorem is that there
exists a well-defined energy gap between the two lowest energy
states of the system. In a more realistic case with decoherence,
however, the energy levels of the qubit register are broadened by
the coupling to environment, as illustrated in Fig.~\ref{fig1}.
Even the simplest environment, e.g., a two-state system, splits a
single anticrossing of the two qubit levels into four
anticrossings with smaller gaps (Fig.~\ref{fig1}b). An environment
with a continuous spectrum turns the anticrossing point into a
continuous region of some width $W$ (Fig.~\ref{fig1}c) within
which incoherent tunneling between the two qubit states can take
place. Thus for such typical models of environment, the gap no
longer exists in the ``qubits+environment'' system. The broadening
$W$ is directly related to the decoherence time of the qubit
states. Any uncertainty $W$ in the energy of an energy eigenstate
makes the accumulated phase of this state also uncertain in time
$\tau_{\rm decoh}\sim 1/W$. Since the broadening $W$ typically
increases with the number of qubits, while the minimum gap $g_m$
decreases, the realistic large-scale system will eventually fall
in the incoherent regime $W\gg g_m$. This means that studies of
the adiabatic theorem do not apply to such {\em realistic}
situations and therefore new ways of understanding AQC performance
become necessary. One possible approach towards this goal is to
generalize the adiabatic theorem to open quantum systems
\cite{Sarandy}.

In this paper, however, we study the evolution of an adiabatic
quantum computer in the ``incoherent'' regime by developing a
corresponding description of Landau-Zener transitions for $W\gg
g_m$. We use the model of decoherence appropriate for solid-state
circuits, where the AQC approach is particularly promising. One
characteristic feature of such a model is that it should allow for
non-Markovian, in particular low-frequency, environmental noise.
Previous studies have mainly considered Markovian environments
\cite{Childs,AKS,we1,TAQC}. A correct description of the
interaction with a low-frequency environment, which has the
strongest effect on the AQC algorithms \cite{TS}, requires a
non-perturbative or strong-coupling theory of the
environment-qubit interaction.

Another feature of our ``solid-state'' approach is the assumption
that the environment responsible for decoherence is in equilibrium
at some temperature $T$, and is sufficiently large to enforce (on
some time scale) the equilibration among the qubit states at the
same temperature. Even the low-frequency noise that dominates the
decoherence of the solid-state qubits (see, e.g., \cite{exp,TLS})
comes usually from equilibrium sources \cite{we2}. Previous studies
of the AQC decoherence used models that do not account directly for
such equilibration \cite{Sarandy,Childs,AKS,roland,AJF}. Since
environment temperature can not be reduced indefinitely, for a
sufficiently large system, $T$ will inevitably be larger than the
minimum gap $g_m$. This means that in contrast to closed systems,
Landau-Zener transitions in the presence of decoherence are
intrinsically linked to thermal excitations out of the ground state,
making it necessary to consider the two types of the transitions
simultaneously.

Quantitatively, we introduce the decoherence as usual by adding the
bath $H_B$ and the interaction Hamiltonian $H_{\rm int}$ to the
Hamiltonian $H_S$ of the qubit register: $H_{\rm total}{=} H_S+H_B+
H_{\rm int}$. As discussed above, we use the two-state approximation
near the anticrossing, assuming that $g_m$ is much smaller than the
energy gaps separating the first two from the other levels
\footnote{Two state approximation is in general valid for the
minimum gaps that result from first order quantum phase transitions.
For the second order quantum phase transitions, other methods become
necessary. See e.g.,  S. Mostame, G. Schaller, and R. Schützhold
Phys. Rev. A {\bf 76}, 030304(R) (2007).}:
\be H_S =-(\epsilon \sigma_z + g_m \sigma_x )/2 , \qquad H_{\rm
int} = - Q \sigma_z/2, \label{H2L} \ee
where $\sigma$'s are the Pauli matrices, $Q$ is an operator of the
environmental noise, $\epsilon \equiv E(s{-}s_m)$ with $E\gg g_m$
defining the energy scale which characterizes the anticrossing at
$s=s_m$. Independent couplings of individual qubits to their
environments produce only the $\sigma_z$-coupled noise in the
two-state model (\ref{H2L}) \cite{we1}. We assume that the noise
is Gaussian so that we do not need to specify $H_B$ explicitly.
Then, all averages can be expressed in terms of the spectral
density:
\[ S(\omega) = \int_{-\infty}^\infty dt \
e^{i\omega t}\langle Q(t)Q(0)\rangle, \]
where $\langle ... \rangle$ denotes averaging over the
environment. Gaussian noise is expected if the environment
consists of a large number of degrees of freedom all weekly
coupled to the system \cite{Leggett}.

In the regime of incoherent Landau-Zener transitions considered in
this work, both the environment-induced broadening $W$ of the two
basis states of the Hamiltonian (\ref{H2L}) and temperature $T$
are taken to be much larger than $g_m$. This means that the time
($\sim 1/W$), during which the two states lose their relative
phase coherence, is much smaller than the typical tunneling time
($\sim 1/g_m$) which implies that the tunneling between these
states will be incoherent. In particular, the off-diagonal
elements of the density matrix $\rho$ of the system (\ref{H2L})
vanish within the time $\tau_{\rm decoh} \sim 1/W$ so that $\rho$
reduces to diagonal elements, i.e. to $\rho_z \equiv p_0-p_1$,
which is governed by the usual kinetic equation
\be \dot \rho_z = -\Gamma (\rho_z-\rho_\infty), \label{kin} \ee
where $\Gamma=\Gamma_{01} + \Gamma_{10}$ and $\rho_\infty =
[\Gamma_{10} - \Gamma_{01}]/\Gamma$. Here we use the standard
notations: $|0\rangle$ and $|1\rangle$ are the two eigenstates of
$\sigma_z$ with eigenvalues $\mp 1$, respectively, $p_j$ is the
occupation probability of state $|j\rangle$, and $\Gamma_{ij}$ is
the rate of tunneling from state $|i\rangle$ to $|j\rangle$.

The physics behind such an incoherent tunneling is the same as for
macroscopic resonant tunneling (MRT) of flux in superconducting
flux qubits which has been studied experimentally \cite{we2} and
theoretically \cite{we3}. In particular, the transition rates have
the structure of resonant peaks of width $W$ in the vicinity of
the anticrossing point. These rates can be explicitly calculated
by a perturbation expansion in $g_m$ and assuming Gaussian noise
\cite{we3}:
\be \Gamma_{01}(\epsilon) = {g_m^2 \over 4} \int dt e^{i\epsilon
t} \exp \left\{ \int {d\omega \over 2\pi} S(\omega)
\frac{e^{-i\omega t}{-}1}{\omega ^2} \right\} . \label{e9} \ee
The rate of the backward tunneling is determined by the relation
$\Gamma_{10}(\epsilon)=\Gamma_{01}(-\epsilon)$. In the case of
white noise, $S(\omega)=S(0)$, Eq.~(\ref{e9}) gives the tunneling
peak in the form of a Lorentzian line-shape:
\be \Gamma_{01}(\epsilon) = {1 \over 2}{g_m^2 W \over
\epsilon^2+W^2}, \qquad W={1\over 2} S(0) \, . \label{Lorentzian}
\ee
On the other hand, in the situation characteristic for practical
solid-state qubits when the noise is dominated by the
low-frequency components, Eq.~(\ref{e9}) reduces to a shifted
Gaussian \cite{we3}:
\ba \Gamma_{01} (\epsilon) = \sqrt{\pi \over 8}{g_m^2 \over W}
\exp \left\{-{(\epsilon - \epsilon_p)^2 \over 2W^2}\right\}, \ \
\label{Gauss} \\
W^2 = \int {d\omega \over 2\pi} S(\omega), \qquad  \epsilon_p =
{\cal P} \int {d\omega \over 2\pi} {S(\omega) \over \omega}.
\nonumber \ea
For environment in thermal equilibrium, the width $W$ and the
position $\epsilon_p$ of the Gaussian are related by \cite{we3}:
\be W^2 = 2T  \epsilon_p \, . \label{equil} \ee
These theoretical results have been experimentally confirmed in
flux qubits \cite{we2}.

\begin{figure}[t]
\includegraphics[width=6.5cm]{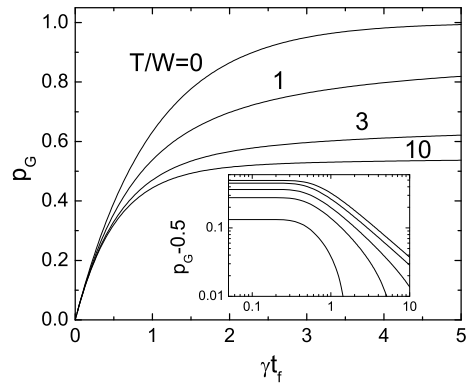}
\caption{The occupation probability $p_G$ of the ground state as a
function of the dimensionless evolution time $\gamma t_f$ for
different temperatures $T$ in the case of the Gaussian tunneling
rates (\ref{Gauss}). The inset shows the dependence of $p_G$ on
$T/W$ for $\gamma t_f \,=\, 1;\,1.5;\,2;\,3;\,5$ from lower to
upper curves respectively. } \label{fig2}
\end{figure}

Let us first study the kinetic equation (\ref{kin}) in two extreme
cases. In the small-$T$ regime $\rho_\infty \simeq \text{sgn}\,
\epsilon$ which implies, with the initial condition $\rho_z(0)=1$,
that the right hand side of (\ref{kin}) is nonzero only for
$\epsilon
> 0$. This leads to the ground state probability
\begin{eqnarray}
& & \;\;\;\;\;\;\;\;\;\;\;\;\;\;\;\;\;\;\; p_G = 1-e^{-\gamma
t_f},
\label{low} \\
& & \gamma \equiv {1\over t_f}\int_0^{\infty} \Gamma(\epsilon)
{d\epsilon \over \dot \epsilon} =  {1\over t_f}
\int_{-\infty}^{\infty} \Gamma_{01}(\epsilon) {d\epsilon \over
\dot \epsilon} \, . \label{Gbar} \end{eqnarray}
We shall see later that under relatively general conditions
$\dot\epsilon \propto 1/t_f$ and therefore $\gamma$ is independent
of $t_f$. These equations assume that the range of $\epsilon$ is
large enough to effectively cover the whole peak of $\Gamma_{01}$,
therefore justifying infinite integration limits. In particular, the
range of $\epsilon$ should be larger than (among other energies) the
cutoff energy of the environment excitations. In the opposite
large-$T$ regime, one has $|\epsilon| \ll T$ and hence $\rho_\infty
= 0$ in Eq.~(\ref{kin}) for energy $\epsilon$ within some relevant
interval around the anticrossing point $\epsilon=0$ (this condition
is made more precise below). The ground stare probability is then
\be p_G = {1\over 2}\left(1-e^{-2\gamma t_f}\right). \label{high}
\ee
Because of the thermal excitations, $p_G$ approaches 1/2 in the
slow-evolution limit. For the intermediate $T$ regime, $p_G$
always falls between (\ref{low}) and (\ref{high}), therefore these
equations give, respectively, upper and lower bounds for the
probability of success (see Fig.~\ref{fig2} and discussion below).

An important feature of (\ref{e9}) is that for uniform evolution,
i.e., $\dot \epsilon =\mbox{const} \equiv \nu$, it gives $\gamma t_f
= {1 \over \nu} \int_{-\infty}^\infty \Gamma_{01} (\epsilon)
d\epsilon = \pi g_m^2/2\nu $, independently of $S(\omega)$, leading
in the small-$T$ regime to the same Landau-Zener probability
(\ref{low}) as in the decoherence-free case. This result extends the
recent proofs \cite{osc,andy,wubs2} that at $T=0$ Landau-Zener
probability is unaffected by decoherence. The physical reason for
this is that the decoherence changes only the profile of the
transition region while keeping the total transition probability the
same. Therefore, in the two extreme regimes, the ground state
probabilities (\ref{low}) and (\ref{high}) are completely
independent of the form of the noise spectrum $S(\omega)$.

At intermediate temperatures, on the other hand, the quantitative
$t_f$-dependence of the probability $p_G$ is sensitive to the
specific form of $S(\omega)$ and therefore to the tunneling rates.
For Gaussian rates (\ref{Gauss}) and uniform evolution, $p_G$
calculated from Eq.~(\ref{kin}) is shown in Fig.~\ref{fig2}. The
curves characterize the transition between the low- (\ref{low}) and
high- (\ref{high}) temperature limits. At small evolution times when
$t_f \ll \gamma^{-1}$ all curves coincide, with $p_G=\gamma t_f$ in
the linear approximation, independently of temperature $T$. The
temperature dependence of $p_G$ appears only in the second-order
terms in $\gamma t_f$. For slow evolution, $t_f \geq \gamma^{-1}$,
$p_G$ varies from 1 to 1/2 with temperature T -- see inset in
Fig.~\ref{fig2}. If the evolution is infinitely slow, the occupation
probabilities of the states $|0\rangle$ and $|1\rangle$ should
always reach the local thermal equilibrium. This, however, is not
the relevant regime for the present discussion. In the relevant
case, the rate $\nu$ is comparable to the maximum tunneling rates
$\Gamma$, and therefore becomes much larger than the tunneling rates
as the system moves away from the resonance, so that the local
equilibrium is not maintained. This means that, strictly speaking,
the large-$T$ result (\ref{high}) is valid for any $t_f$ only for
$T\gg E$. Asymptotic analysis of the evolution equation for the case
of the Gaussian rates (\ref{Gauss}) shows that in the more
interesting regime when $T\gg W$ but $T\ll E$, the ground-state
probability is:
\be p_G = \frac{1}{2} + \frac{W}{\sqrt{2} T}[\ln \gamma t_f
]^{1/2}. \label{as} \ee
This equation describes the increase of $p_G$ towards the local
equilibrium at sufficiently large evolution time $t_f$, and
corresponds to the large-$T$ part of the two curves with larger
$\gamma t_f$ in the inset in Fig.~\ref{fig2}.

We now use the results presented above to discuss the performance
of AQC in the incoherent regime $g_m \ll W, T$. For this, one
needs to distinguish global and local adiabatic evolutions. In the
{\em global} scheme, the adiabatic evolution is uniform, $\dot
\epsilon = \mbox{const}=E/t_f$, and Eqs.~(\ref{low}) and
(\ref{high}) show that the required computational time $t_f \simeq
\gamma^{-1} = 2E/\pi g_m^2$ coincides with the decoherence-free
case independently of decoherence and temperature $T$. Even if the
large $T$ reduces $p_G$ to $\simeq 1/2$, to find correct solution,
one only needs to repeat the computation process on average two
times.

Global adiabatic evolution, however, does not yield the optimal
performance in coherent AQC. Indeed, for the case of adiabatic
Grover search \cite{roland}, the global adiabatic scheme yields
the complexity of the classical exhaustive search, i.e., $t_f =
O(N)$, where $N$ ($=2^n$) is the size of data base. In the more
efficient {\em local} scheme \cite{roland}, one takes $\dot
\epsilon(t) = \alpha g(t)^2$, so that the adiabatic condition is
satisfied uniformly (the system slows down in the region of small
gap) and the computation time is $t_f =\pi/\alpha g_m$ which for
the case of adiabatic Grover search yields the optimal
$O(\sqrt{N})$ performance.
The local evolution plays crucial role for the scaling analysis of
the AQC \cite{roland,farhi05,mosca07}, although in some cases it
is only assumed implicitly. In general, however, finding the gap
$g(s)$ is as hard as solving the original problem, and only in
some cases, e.g, the adiabatic Grover search, $g(s)$ is
independent of the final solution and can be found {\em a priori}
analytically.

The enhanced performance of the local scheme comes at a price of
its stronger sensitivity to decoherence. A qualitative reason for
sensitivity of local AQC is that although decoherence does not
change the total integral transition probability, it distributes
it over a much larger energy interval $W\gg g_m$, making it
necessary to slow down the evolution for a longer period of time.
If one uses the same $\epsilon(t)$ as in the decoherence-free
case, the average tunneling rate (\ref{Gbar}) is dominated by the
vicinity of the point $\epsilon =0$. Quantitatively, $\dot\epsilon
= \alpha g^2$ and $t_f =\pi/\alpha g_m$ yield
$(t_f\dot\epsilon)^{-1} = g_m/\pi g^2 \approx \delta (\epsilon)$,
which together with (\ref{Gbar}) and (\ref{e9}) give $\gamma
\approx \Gamma_{01}(0) \propto g_m^2$. Therefore the computation
time is $t_f \simeq \gamma^{-1} \propto g_m^{-2}$, which is
similar to the performance of the global scheme with the only
possible enhancement compared to the global case being a
prefactor. In the case of white noise, Eq.~(\ref{Lorentzian})
leads to $\gamma = g_m^2/2W$, while for the low-frequency noise,
Eq.~(\ref{Gauss}) gives $\gamma = \sqrt{\pi /8} (g_m^2 /W)
e^{-W^2/8T^2}$. Notice that in the latter case, lowering $T$ with
constant width $W$ \cite{we2} does not shorten the computation
time.

To summarize, we have studied the decoherence effects on AQC due to
general non-Markovian environments in the strong decoherence regime
in which the broadening of the energy levels completely smears out
the anticrossing region. Our strong coupling treatment shows that
global AQC remains unaffected by strong decoherence $W>g_m$ and is
independent of the type of noise, while the local AQC provides only
a prefactor improvement of the algorithm running time in this regime
and does not change the scaling of this time with $g_m$ as compared
to the case without decoherence. Thus, the local AQC can only
maintain its properties if $W<g_m$. Since $W \sim 1/\tau_{\rm
decoh}$, and $t_f \sim 1/g_m$ for the local scheme in the
weak-decoherence regime, the computation time is limited by the
decoherence $t_f<\tau_{\rm decoh}$ in the same way as in gate model
QC. Therefore, the advantageous scaling of the local AQC requires
phase coherence throughout the evolution as in the gate model.
Insensitivity of AQC to decoherence only holds for the global scheme
and does not apply to local AQC. It should be emphasized that in our
treatment we have assumed that the minimum gap is a result of a
first order quantum phase transition for which two-state model holds
and the broadening of the energy levels and also thermal excitation
do not mix the lowest two states with other excited states. For
stronger noise or higher temperatures, one needs to take higher
states into consideration.

The authors are thankful to A.J.~Berkley, J.D.~Biamonte, V.~Choi,
R.~Harris, J.~Johansson, M.W.~Johnson, W.M.~Kaminsky, T.~Lanting,
D.A.~Lidar, and G.~Rose for fruitful discussions.

\end{document}